# Resonance Measurement of Nonlocal Spin Torque in a Three-Terminal Magnetic Device

Lin Xue[1], Chen Wang[1], Yong-Tao Cui[1], Luqiao Liu[1], A. Swander[1], J. Z. Sun[3], R. A. Buhrman[1] and D. C. Ralph[1,2]

[1]Cornell University and [2]Kavli Institute at Cornell, Ithaca, NY 14853 USA

[3]IBM T. J. Watson Research Center, Yorktown Heights, New York 10598, USA

**Abstract**

A pure spin current generated within a nonlocal spin valve can exert a spin transfer torque on a nanomagnet. This nonlocal torque enables new design schemes for magnetic memory devices that do not require the application of large voltages across tunnel barriers that can suffer electrical breakdown. Here we report a quantitative measurement of this nonlocal spin torque using spin-torque-driven ferromagnetic resonance. Our measurement agrees well with the prediction of an effective circuit model for spin transport. Based on this model, we suggest strategies for optimizing the strength of nonlocal torque.

Spin transfer torque enables the efficient manipulation of magnetization in nanoscale magnetic devices [1-3]. Spin torque due to the flow of a spin-polarized charge current within conventional two-terminal magnetic tunnel junctions (MTJs) and magnetic multilayer devices has been studied intensively and is being developed for technology. In addition, it has been shown recently that in multiterminal device structures a spin torque can also be exerted by a *nonlocal* pure spin current (meaning a spin current associated with zero net charge flow, as distinct from a spin-polarized charge current) [4-7], in agreement with predictions [8]. This nonlocal spin torque can be sufficiently strong to cause magnetic reversal [4-7]. However, thus far the only means of detecting nonlocal spin torques in multiterminal devices has been to observe full magnetic reversal, which does not provide a quantitative torque measurement and which yields information only in the high bias regime. Here we report measurements of nonlocal spin torque using spin-torque-driven ferromagnetic resonance (ST-FMR) [9-14], a technique that is both quantitative and that operates for any applied bias. We compare the measured nonlocal torque to the prediction of an effective circuit model of spin transport, finding reasonable agreement, and we suggest strategies for further optimization.

The device geometry we consider is a three-terminal structure consisting of a lower all-metal spin valve with a MTJ on top [Fig. 1(a)]. Nonlocal spin-torque switching has been measured previously by the IBM group in devices with the same design, except for a slightly thicker spin injection layer [6]. An applied charge current passes from a bottom TaN electrode (terminal T1) approximately 100 nm in diameter through an exchange-biased PtMn(17.5 nm)/$Co_{70}Fe_{30}$(3.5 nm) bilayer (magnetic layer F1) and out of the device laterally through a PtMn(17.5 nm)/$Co_{70}Fe_{30}$(3.5

nm)/Cu(N)(30 nm) multilayer (terminal T2), where Cu(N) means nitrogen-doped Cu. This generates spin accumulation in the Cu(N) channel above the TaN contact. A pure spin current can then diffuse to a 2 nm $Co_{60}Fe_{20}B_{20}$ layer (magnetic layer F2) positioned above the Cu(N) channel. This layer F2 will serve as the magnetic free layer in the experiment, reorienting in response to the nonlocal spin torque. The cross section of F2 is approximately an ellipse, 70 × 150 $nm^2$, with the long axis parallel to the exchange bias direction of F1. We have also measured 80 × 120 $nm^2$ and 90 × 200 $nm^2$ devices with similar results. The device structure is completed by an MgO-based MTJ positioned above F2, whose magnetoresistance (measured between terminals T2 and T3) depends on the orientation of F2. We will discuss data for a sample with a MTJ resistance of 30.9 kΩ in the parallel magnetic state with a tunneling magnetoresistance of 39%, and with a metallic channel resistance (between the contact pads of terminals T1 and T2) of 23 Ω.

To perform an ST-FMR measurement of the nonlocal spin torque, we first apply a magnetic field $H$ in the sample plane approximately perpendicular to the exchange-bias direction so as to turn the magnetization of the free layer F2 away from the magnetization of F1 and F3. Layer F2 has a small coercive field (~ 30 Oe), so that to a good approximation in a magnetic field of order 1 kOe it aligns to the field direction. Layers F3 and F1 are reoriented by lesser amounts because F3 is part of a synthetic antiferromagnet and F1 is subject to an approximately 1.1 kOe exchange bias through interaction with PtMn (see Figs. 1(b,c)). The next step of the measurement is to apply a pulsed microwave-frequency current with magnitude $I_{RF}^{applied}$ between the contact pads leading to terminals T1 and T2. This produces an oscillatory nonlocal spin torque that causes the magnetization of the free layer to precess. We measure the precession by detecting a dc voltage that results across the MTJ (between terminals T2 and T3) as a consequence of mixing between the oscillating resistance of the MTJ and an oscillating current $I_{RF}^{leakage}$ of order $10^{-3} I_{RF}^{applied}$ that flows through the MTJ. (If $I_{RF}^{leakage}$ had been too small to provide a mixing measurement of the resonance, we could also have applied a separate microwave current directly to the MTJ to give the same effect.) All measurements are performed at room temperature, and we use the convention that negative currents correspond to electron flow in the direction of the arrows in Fig. 1(a) (giving a torque favoring parallel alignment between F2 and F1).

Figure 1(d) shows an example of a nonlocal ST-FMR resonance peak measured for a fixed microwave frequency $\omega/(2\pi)$ = 12 GHz, for a swept magnetic field oriented 75° from the exchange bias direction of layer F1 and for a dc current $I_{dc}^{SV}$ = 5 mA applied between terminals T1 and T2. We used excitation currents $I_{RF}^{applied}$ < 1.9 mA, and verified that the output mixing signal scaled $\propto (I_{RF}^{applied})^2$ so that the magnetic response is in the linear regime.

The lineshapes of the nonlocal ST-FMR signals can be understood by modeling the dynamics of the magnetic free layer in a macrospin approximation and adapting the theory used to analyze ST-FMR in a two-terminal MTJ [14], with the result that the resonant part of the signal should have the simple form [15]:

$$\text{Resonance} \propto c_S S(\omega, H) + c_A A(\omega, H). \tag{1}$$

Here $S(\omega, H) = \left[1 + \left(\omega - \omega_m(H)\right)^2 / \sigma^2\right]^{-1} \approx \left[1 + (H - H_m)^2 / (\Delta H)^2\right]^{-1}$ is a symmetric Lorentzian peak as a function of $\omega$ or $H$, $A(\omega, H) = \left[(\omega - \omega_m(H))/\sigma\right] S(\omega, H)$ is an antisymmetric

Lorentzian with the same linewidth, $\omega_m$ is the resonance frequency at a given value of $H$ [15], $\sigma$ is the frequency linewidth, $H_m$ is the resonance field at a given value of $\omega$, and $\Delta H \approx \sigma / [d\omega_m / dH]$ is the field linewidth. The prefactors $c_S$ and $c_A$ are to a good approximation constant as a function of $H$ in the region of the resonance, but they depend on the current and $\omega$. The measurement may also contain a nonresonant background that can depend weakly on $H$. The linewidth parameter $\sigma$ is predicted [15] to depend on the magnitude of the in-plane component $\tau_\parallel$ of the spin transfer torque in the form

$$\sigma \approx \frac{\alpha \gamma M_{eff}(N_x + N_y)}{2} - \frac{\gamma}{M_s Vol} \left. \frac{\partial \tau_\parallel (I_{SV}, \theta_{SV})}{\partial \theta_{SV}} \right|_{I_{SV}}. \qquad (2)$$

Here $\alpha$ is the Gilbert damping coefficient, $\gamma = 2\mu_B / \hbar$ is the absolute value of the gyromagnetic ratio, $4\pi M_{eff}$ is the effective in-plane anisotropy of layer F2, $N_x = 4\pi + H / M_{eff}$, $N_y = H / M_{eff}$, $M_s Vol$ is the total magnetic moment of F2; $I_{SV}$ is the current in the spin-valve part of the device between terminals T1 and T2, and $\theta_{SV}$ is the offset angle between F2 and F1. For an all-metal spin valve, the spin torque should have only an in-plane component (*i.e.*, in the direction $\hat{m} \times (\hat{m} \times \hat{M}) / |\hat{m} \times \hat{M}|$, where $\hat{m}$ is the orientation of the free layer moment and $\hat{M}$ is the orientation of the polarizer layer) [3], so Eq. (2) allows a measurement of the full nonlocal spin transfer torque.

A fit of Eq. (1) to a measured resonance lineshape is included in Fig. 1(d), using the fitting parameters $\sigma = (5.94 \pm 0.08) \times 10^8$ rad·Hz, and $c_S / c_A = -1.33 \pm 0.03$. We allow for a linear dependence on $H$ for the nonresonant background, but ignore the weak dependence of $\theta_{SV}$ and $\sigma$ on $H$ near the resonance. The fit in Fig. 1(d) is excellent, and we observe a similar quality of agreement for different values of $\omega$, field angle, and $I_{SV}$. From the measured resonance frequencies we determine $4\pi M_{eff} = 13 \pm 1$ kOe [15].

The strength of the nonlocal spin torque can be determined most accurately [15] from the resonance measurement by using Eq. (2) to analyze the dependence of the resonance linewidth on $I_{SV}$. A similar approach has been used previously to measure spin torque in magnetic tunnel junctions [16] and due to the spin Hall effect [17,18]. We show in Fig. 2(a) the measured evolution of the resonance as a function of $I_{dc}^{SV}$ (the dc component of $I_{SV}$), for $\omega/(2\pi) = 12$ GHz and a field orientation 75° relative to the exchange bias direction. We observe that the linewidth depends linearly on $I_{dc}^{SV}$ [Fig. 2(b)]. By fitting to Eq. (2) and using as above that $4\pi M_{eff} = 13 \pm 1$ kOe (with $M_S = 1100$ emu/cm$^3$ [11] and with the free-layer volume $Vol = 1.7 \times 10^{-17}$ cm$^3$), we determine $\partial \tau_\| / \partial I_{SV}\big|_{\theta_{SV}} = 0.05 \pm 0.01$ $(\hbar/2e)$ and $\alpha = 0.012 \pm 0.002$ for these experimental conditions.

We have carried out similar measurements of linewidth versus $I_{dc}^{SV}$ for field angles of 60° and 75° and for field magnitudes yielding resonance frequencies from 8 to 12 GHz. When comparing results for different fields, we take into account that the nonlocal spin torque should be proportional to the component of the spin current perpendicular to the free layer magnetization, so that $\tau_\| = (\hbar/2e)\eta_\| I_{SV} \sin\theta_{SV}$ (or $\partial\tau_\|/\partial\theta_{SV}\big|_{I_{SV}} = (\hbar/2e)\eta_\| I_{SV} \cos\theta_{SV}$), where $\eta_\|$ is a dimensionless efficiency. We estimate $\theta_{SV}$ by assuming that the magnetization of F2 aligns with the applied field and calculating the magnetization angle of F1 by assuming that it responds as a macrospin to the combined action of $H$ and the exchange field $H_{ex} = 1.1 \pm 0.2$ kOe [19].

Figure 2(c) shows separate measurements of the spin-torque efficiency $\eta_\parallel$ for a range of field magnitudes (0.6 - 1.3 kOe at an angle of 75°), that correspond to resonance frequencies of 8-12 GHz and offset angles $\theta_{SV}$ between 49° and 35°. Our final overall value for the efficiency of the nonlocal spin torque is $\eta_\parallel = 0.10 \pm 0.02$.

Sun et al. [6] performed spin-torque switching experiments with devices of the same structure except with a slightly thicker injection layer F1, and obtained a zero temperature critical switching current $I_{c0} = -6.84$ mA for $\theta_{SV}$ near 180° and $I_{c0} = 7.20$ mA for $\theta_{SV}$ near 0° for a device cross section of 69 × 161 nm². For an in-plane magnetized free layer in zero external field, $I_{c0} \approx (2e/\hbar)(\alpha M_S Vol / \eta_\parallel)$ [20]. Therefore the switching measurement can also be used to estimate the spin torque efficiency $\eta_\parallel$ if $\alpha$ and $4\pi M_{eff}$ are known. Using the values obtained above from our resonance measurements, $\alpha = 0.012 \pm 0.002$ and $4\pi M_{eff} = 13 \pm 1$ kOe, the switching currents from ref. [6] correspond to an in-plane spin-torque efficiency $\eta_\parallel = 0.07 \pm 0.02$, consistent with our ST-FMR result.

The value of the nonlocal torque that should be expected theoretically can be estimated using an effective circuit model [21-24] for spin transport. For the case $\theta_{SV} = 90°$, the simple effective circuit in Fig. 3 applies. (For other angles, as noted above, we expect the spin torque should be proportional to $\sin\theta_{SV}$.) In this circuit model, we assume that the spin accumulation relaxes only by flow to the free layer F2 or by flow through the normal contact N' toward T2. Using materials parameters appropriate to our sample geometry, we estimate that the spin-dependent resistances appropriate for N', the spin injector layer F1, the Cu(N) spacer N, and the free layer F2 are approximately $R^{N'} \approx 0.6 \pm 0.2$ Ω, $R^{F1}_\uparrow \approx 0.07$ Ω, $R^{F1}_\downarrow \approx 0.29 \pm 0.08$ Ω, $R^N \approx 0.44$ Ω, and $R^{F2}_\perp \approx 0.016$ Ω [15]. Solving the circuit, the calculated spin torque efficiency is

$$\eta_{\text{circuit}} \equiv \frac{2I_S}{I_{SV}} = \frac{R^{N'}\left(R^{F1}_\downarrow - R^{F1}_\uparrow\right)}{\left(R^{F1}_\downarrow + R^{F1}_\uparrow\right)\left(R^{F2}_\perp + R^N + R^{N'}\right) + 2R^{N'}\left(R^{F2}_\perp + R^N\right)} \approx 0.14 \pm 0.04. \quad (3)$$

The prediction of the circuit model is therefore in quite reasonable agreement with our measurement.

To achieve optimal efficiency based on Eq. (3), the device parameters should satisfy three conditions: (i) a large intrinsic injector polarization $P = \left(R^{F1}_\downarrow - R^{F1}_\uparrow\right)/\left(R^{F1}_\downarrow + R^{F1}_\uparrow\right)$, (ii) a small spin resistance for electrons going from the injector to the magnetic free layer to apply a spin torque, $R^{F2}_\perp + R^N \ll R^{F1}_\downarrow + R^{F1}_\uparrow$, and (iii) a large spin resistance for electrons flowing toward terminal T2, $R^{N'} \gg R^{F2}_\perp + R^N$, so as to prevent spin current from escaping by this path rather than applying a torque to F2. However, in the existing device design, neither conditions (ii) or (iii) are fully satisfied. To improve the spin torque efficiency, the effective resistance of the spin injector (layer F1) can be increased relative to $R^N$, perhaps by using tunnel-barrier injection, by decreasing the thickness of the Cu(N) layer below 30 nm, and/or by reducing the resistivity of the Cu(N) layer. The device performance can also be improved by increasing $R^{N'}$ relative to $R^N$ by reducing the thickness of the 30 nm Cu(N) layer and/or by increasing the spin relaxation length $l^{N'}_{SF}$ by eliminating the PtMn/Co$_{70}$Fe$_{30}$ layers underneath the portion of the Cu(N) layer not adjacent to the injector region. If conditions (ii) and (iii) are fully met, then the optimum nonlocal spin torque efficiency should equal the injector polarization, $\eta_{\text{circuit}} = P$, meaning that the nonlocal spin torque can be made just as efficient as the spin torque in conventional two-terminal devices.

In summary, we have performed an ST-FMR measurement of the nonlocal spin torque due to a pure spin current in a three-terminal device. We measure a spin torque efficiency $\partial \tau_\parallel / \partial I_{SV} \left[ 2e / (\hbar \sin\theta_{SV}) \right] = 0.10 \pm 0.02$. This agrees well with the efficiency expected within an effective circuit model. Based on the circuit analysis, we estimate that the nonlocal device geometry can be optimized so that the strength of the nonlocal torque should reach $\partial \tau_\parallel / \partial I_{SV} = P \sin\theta_{SV} \, \hbar / 2e$, the same value expected for the local spin torque in two-terminal devices. Due to the low resistance of the spin-valve current channel in the three-terminal devices, the ratio of the spin torque to the applied power is already much greater in the existing three-terminal devices than in two-terminal MTJs. The nonlocal spin torque in three-terminal devices therefore possesses a combination of virtues relative to conventional MTJs -- reduced susceptibility to tunnel barrier breakdown and reduced power consumption together with high spin torque efficiency -- that can make this device geometry an interesting candidate for applications.

We thank Erich Mueller and Bo Xiang for helpful discussions. Cornell acknowledges support from ARO, NSF (DMR-1010768), and the NSF/NSEC program through the Cornell Center for Nanoscale Systems. We also acknowledge NSF support through use of the Cornell Nanofabrication Facility/NNIN and the Cornell Center for Materials Research facilities (DMR-1120296).

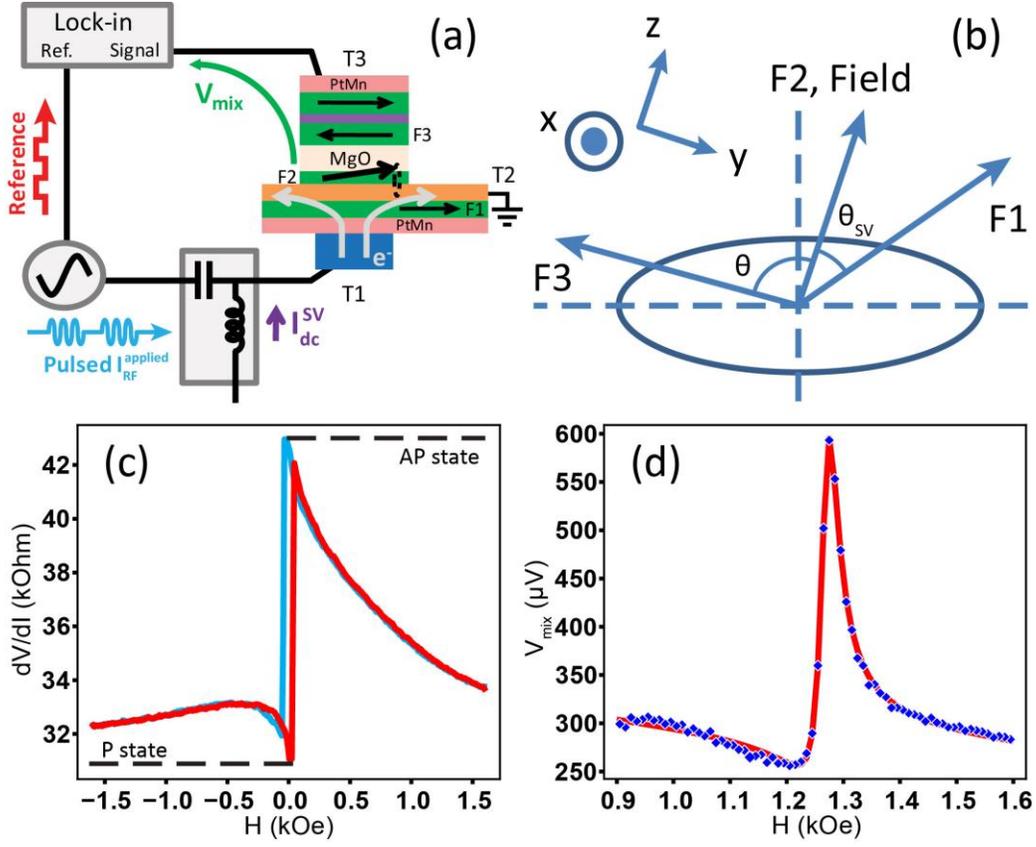

FIG. 1. (a) Illustration of the ST-FMR circuit. (b) Orientations of the magnetic moments of layers F1, F2, and F3 when a magnetic field of 1.3 kOe is applied 75 ° from the exchange bias direction. (c) Differential resistance vs. external magnetic field applied 75 ° from the exchange bias direction. The resistances for parallel and antiparallel alignment between F2 and F3 are indicated. (d) (points) Measured ST-FMR signal at 12 GHz for a magnetic field orientation 75 ° from the exchange bias direction ($\theta_{SV} \approx 35°$ at resonance) with $I_{dc}^{SV} = 5$ mA. (line) Fit to Eq. (1) assuming a linear dependence on $H$ for the background.

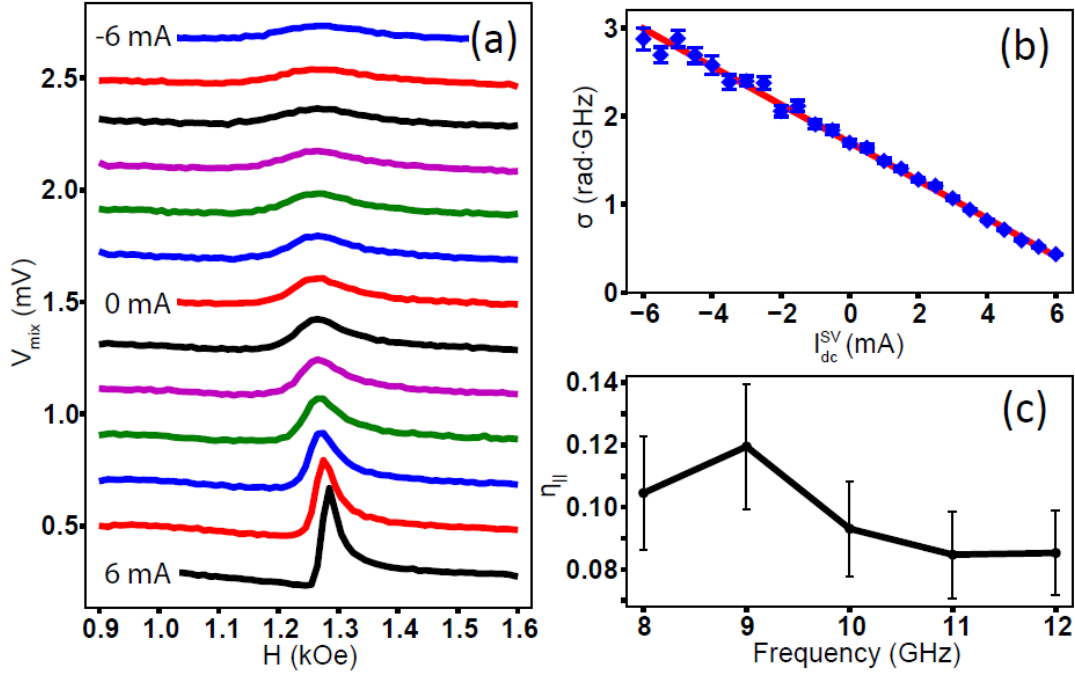

FIG. 2. (a) ST-FMR signals measured, for different values of $I_{dc}^{SV}$, at 12 GHz for a magnetic field orientation 75° from the exchange bias direction ($\theta_{SV} \approx 35°$ at resonance). Curves are offset vertically by 0.2 mV. (b) Dependence of resonant linewidth $\sigma$ on $I_{dc}^{SV}$ for the data in (a). (c) Efficiency of the in-plane spin torque, defined as $\eta_\| = \left[2e/(\hbar I_{SV} \cos\theta_{SV})\right] \partial \tau_\| / \partial \theta_{SV} \big|_{I_{SV}}$, determined from ST-FMR measurements of $\sigma$ vs. $I_{dc}^{SV}$ together with Eq. (2), for different values of resonant microwave frequency.

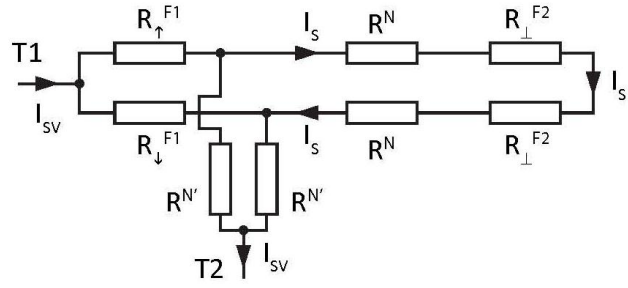

FIG. 3. Effective circuit for modeling spin currents when $\theta_{SV} = 90°$. The total current of spin angular momentum absorbed by layer F2 is $2(\hbar/2e)I_S$.

# Supplement to
# "Resonance Measurement of Nonlocal Spin Torque in a Three-Terminal Magnetic Device"

Lin Xue[1], Chen Wang[1], Yong-Tao Cui[1], Luqiao Liu[1], A. Swander[1], J. Z. Sun[3], R. A. Buhrman[1] and D. C. Ralph[1,2]

[1]Cornell University and [2]Kavli Institute at Cornell, Ithaca, NY 14853 USA

[3]IBM T. J. Watson Research Center, Yorktown Heights, New York 10598, USA

## 1. Derivation of the Resonance Line Shape For Nonlocal ST-FMR

To derive the line shapes of the nonlocal ST-FMR signals we adapt the theory used to analyze ST-FMR in a 2-terminal MTJ [S1]. We assume the free layer is uniformly magnetized in the direction $\hat{m}(t)$, while the other magnetic layers remain fixed in response to $I_{RF}^{applied}$. We use the convention that the $z$ axis lies in the sample plane along the equilibrium direction of $\hat{m}$ in the absence of any applied microwave current, with the $x$ and $y$ axes as shown in Fig. 1(b) of the main paper. We assume that all of the magnetic layers have equilibrium orientations in the sample plane, and that the dynamics of $\hat{m}(t)$ are governed by the Landau-Lifshitz-Gilbert-Slonczewski equation of motion:

$$\frac{d\hat{m}}{dt} = -\gamma \hat{m} \times \vec{H}_{\text{eff}} + \alpha \hat{m} \times \frac{d\hat{m}}{dt} + \gamma \frac{\tau_{\parallel}(I_{SV}, \theta_{SV})}{M_s Vol} \hat{m} \times \frac{\hat{m} \times \hat{M}}{|\hat{m} \times \hat{M}|} + \gamma \frac{\tau_{\perp}(I_{SV}, \theta_{SV})}{M_s Vol} \frac{\hat{m} \times \hat{M}}{|\hat{m} \times \hat{M}|}. \quad (S1)$$

Here $\hat{M}$ is the orientation of layer F1; $M_s Vol$ is the total magnetic moment of F2; $\theta_{SV}$ is the offset angle between F2 and F1; $I_{SV} = I_{RF}^{SV} + I_{dc}^{SV}$ is the charge current between terminals T1 and T2; $\tau_{\parallel}(I_{SV}, \theta_{SV})$ and $\tau_{\perp}(I_{SV}, \theta_{SV})$ are the in-plane and perpendicular components of the current-induced torque; $\vec{H}_{eff} = -N_x M_{eff} m_x \hat{x} - N_y M_{eff} m_y \hat{y}$ is the effective field acting on F2, with $N_x = 4\pi + H/M_{eff}$, $N_y = H/M_{eff}$, and $H$ the component of external field along $\hat{z}$; $\gamma = 2\mu_B/\hbar$ is the absolute value of the gyromagnetic ratio; and $\alpha$ is the Gilbert damping coefficient. We expect that the spin torque within a nonlocal all-metal spin valve should have only an in-plane component [S2], but we include an out-of-plane component as well to account for possible torques due to current-induced Oersted fields. The unit vectors $\hat{m} \times \hat{M}/|\hat{m} \times \hat{M}|$ and $\hat{m} \times (\hat{m} \times \hat{M})/|\hat{m} \times \hat{M}|$ in Eq. (S1) do not possess any dependence on the magnetic field as long as it is applied in the sample plane.

By solving Eq. (S1) for a small enough $I_{RF}^{applied}$ that the magnetic response is in the linear regime we find

$$m_y = \frac{\gamma I_{RF}^{SV}}{2M_s Vol} \frac{1}{\omega - \omega_m - i\sigma} \left( i \frac{\partial \tau_{\parallel}(I_{dc}^{SV}, \theta_{SV})}{\partial I_{SV}} \bigg|_{\theta_{SV}} + \frac{\gamma N_x M_{eff}}{\omega} \frac{\partial \tau_{\perp}(I_{dc}^{SV}, \theta_{SV})}{\partial I_{SV}} \bigg|_{\theta_{SV}} \right), \quad (S2)$$

where

$$\omega_m \approx \gamma M_{eff} \sqrt{N_x \left( N_y - \frac{1}{M_{eff} M_s Vol} \frac{\partial \tau_{\perp}(I_{dc}^{SV}, \theta_{SV})}{\partial \theta_{SV}} \bigg|_{I_{SV}} \right)}, \quad (S3)$$

$$\sigma \approx \frac{\alpha \gamma M_{eff}(N_x + N_y)}{2} - \frac{\gamma}{M_s Vol} \frac{\partial \tau_{\parallel}(I_{dc}^{SV}, \theta_{SV})}{\partial \theta_{SV}} \bigg|_{I_{SV}}. \quad (S4)$$

When the magnetization of F2 precesses, the angle between the magnetizations of F2 and F3

changes as $\delta\theta_{TJ} = \text{Re}(m_y)$. We write that the current through the MTJ can be separated into oscillatory and dc parts as $I_{TJ}(t) = \delta I_{TJ}(t) + I_{dc}^{TJ}$ where $\delta I_{TJ}(t) = \text{Re}[I_{RF}^{leakage}e^{i\phi+i\omega t}]$ and the phase $\phi$ is defined relative to $I_{RF}^{SV}$. (Because of parasitic capacitances and/or inductances in the measurement circuit, this relative phase $\phi$ between the microwave leakage current through the tunnel junction $\delta I_{TJ}(t)$ and the microwave current through the spin valve $I_{RF}^{SV}$ is in general non-zero.) The dc mixing voltage across the MTJ then takes the form

$$V_{mix} \approx \frac{1}{4}\frac{\partial^2 V}{\partial I_{TJ}^2}\left(I_{RF}^{leakage}\right)^2 + \frac{1}{2}\frac{\partial^2 V}{\partial I_{TJ}\partial\theta_{TJ}}\frac{\hbar\gamma\sin\theta_{SV}}{4eM_sVol\sigma}I_{RF}^{leakage}I_{RF}^{SV}\left[c_S S(\omega,H)+c_A A(\omega,H)\right] \quad (S5)$$

with $c_S = -\xi_\parallel\cos\phi + \xi_\perp\Omega_\perp\sin\phi$, $c_A = \xi_\perp\Omega_\perp\cos\phi + \xi_\parallel\sin\phi$,

$S(\omega,H) = \left[1+\left(\omega-\omega_m(H)\right)^2/\sigma^2\right]^{-1} \approx \left[1+(H-H_m)^2\left(d\omega_m(H_m)/dH\right)^2/\sigma^2\right]^{-1}$,

$A(\omega,H) = \left[(\omega-\omega_m(H))/\sigma\right]S(\omega,H)$, and $H_m$ being the field value for which $\omega_m = \omega$. Here $\xi_\parallel = \left[2e/(\hbar\sin\theta_{SV})\right]d\tau_\parallel/dI_{SV}$ and $\xi_\perp = \left[2e/(\hbar\sin\theta_{SV})\right]d\tau_\perp/dI_{SV}$ represent the in-plane and perpendicular torkances in dimensionless units and $\Omega_\perp = \gamma N_x M_{eff}/\omega_m$. As we noted in the main text, the resonant lineshape [the final term in Eq. (5)] is a weighted sum of $S(\omega,H)$ (a symmetric Lorentzian peak as a function of $\omega$ or H) and $A(\omega,H)$ (an antisymmetric Lorentzian ), both with the same frequency linewidth $\sigma$. The first term in Eq. (5) is a nonresonant background that may depend weakly on H.

In order to obtain experimental data to fit to Eq. (S5), we performed the ST-FMR measurements at fixed microwave frequency while sweeping magnetic field, rather than at fixed field while sweeping frequency as in our previous experiments [S1,S3-S5], because in a 3-terminal device it is not possible to calibrate the externally-applied microwave current so that both $I_{RF}^{SV}$ and $I_{RF}^{leakage}$ are simultaneously kept flat as the frequency is varied. In our fitting, we allow for a linear dependence on *H* for the nonresonant background in Eq. (S5), but we assume that $I_{RF}^{SV}$ and $I_{RF}^{leakage}$ are independent of *H* and we ignore the weak dependence of $\theta_{SV}$, $\theta_{TJ}$, and $\sigma$ on *H* near the resonance.

**2. Why Do We Determine The Spin Torque Magnitude From the Resonance Linewidth Rather Than The Resonance Amplitude?**

In principle, one can also measure the spin torque by analyzing the amplitude of the ST-FMR resonance peak (see Eq. (S5) above), to reach a separate determination that is independent of our analysis of the resonance linewidth. We have used analyses of resonance amplitudes previously to measure the magnitude of the local spin torque in 2-terminal magnetic tunnel junctions [S4,S5]. However, because the 3-terminal devices we analyze here were not designed with microwave-frequency experiments in mind, parasitic capacitances in the device structure produce current shunting and a non-zero phase shift $\phi$ between $I_{RF}^{TJ}$ and $I_{RF}^{SV}$ that is difficult to quantify accurately. This adds significant uncertainty to an analysis of the resonance amplitudes because it can alter the relative amplitudes of the symmetric and antisymmetric parts of the resonance (see Eq. (S5) and the definitions for $c_S$ and $c_A$ that follow it). For determining the magnitude of the spin torque in our 3-terminal devices, it is therefore more accurate to analyze the current dependence of the resonance linewidth, because this quantity has no dependence on $\phi$.

## 3. Estimation of Sample Parameters for the Effective Circuit Model of Spin Transport

We estimate the spin resistances for the effective circuit drawn in Fig. 3 of the main paper as follows:

The spin resistances of the $Co_{70}Fe_{30}$ spin polarizing layer F1 are estimated as

$R_{\uparrow}^{F1} = 2\rho_{F1} t_{F1} / \left[ A_{F1}(1+P) \right]$ and $R_{\downarrow}^{F1} = 2\rho_{F1} t_{F1} / \left[ A_{F1}(1-P) \right]$, with $\rho_{F1} = 130$ nΩ·m [S6] the resistivity of this layer, $t_{F1} = 3.5$ nm its thickness, $A_{F1} = 7900$ nm$^2$ its area, and $P = \left( R_{\downarrow}^{F1} - R_{\uparrow}^{F1} \right) / \left( R_{\downarrow}^{F1} + R_{\uparrow}^{F1} \right) = 0.6 \pm 0.1$ [S6,S7] its spin polarization. Here we ignore any additional spin-dependent interface resistance because its effects should be negligible [S8]. We also make the rough approximation that $t_{F1} \approx l_{SF}^{F1}$, the spin diffusion length in CoFe. This has been measured to be $l_{SF}^{F1} \approx 9$ nm at 4.2 K and should be smaller at room temperature [S8].

For the spin resistance of the Cu(N) spacer layer we use $R^N = 2\rho_N t_N / A_N$, with $\rho_N = 60$ nΩ·m [S9], $t_N = 30$ nm, and $A_N = 8200$ nm$^2$.

For the spin resistance of that part of the $Co_{60}Fe_{20}B_{20}$ free layer F2 over which an incident spin component with orientation perpendicular to the magnetization of F2 will be reoriented by precession, we estimate $R_{\perp}^{F2} = 2\rho_{F2} l_{ex} / A_{F2}$, with $\rho_{F2} = 130$ nΩ·m [S6] and $A_{F2} = 8200$ nm$^2$. The precession length should be of order the lattice spacing, $l_{ex} \sim 0.5$ nm [S2].

Finally, for the effective spin resistance corresponding to radial diffusion away from the circular contact toward terminal T2, we estimate $R^{N'} = 2\rho_{N'} \ln\left[ \left( r_{eff} + l_{SF}^{N'} \right) / r_{eff} \right] / (2\pi t_{N'})$. Here $\rho_{N'} = 60$ nΩ·m [S9], $r_{eff} \approx 25$ nm is half the injector radius, and $t_{N'} \approx 30$ nm. $l_{SF}^{N'}$ is an effective spin diffusion length for the $Co_{70}Fe_{30}$(3.5 nm)/Cu(N)(30 nm) electrode that we estimate to be $l_{SF}^{N'} \approx$ 34 nm $\pm 50\%$ based on a layer-thickness-weighted average of the relaxation rates in $Co_{70}Fe_{30}$ and Cu [S10].

Our final estimates for the resistances in the effective circuit are $R_\uparrow^{F1} \approx 0.07$ Ω, $R_\downarrow^{F1} \approx 0.29 \pm 0.08$ Ω, $R^N \approx 0.44$ Ω, $R_\perp^{F2} \approx 0.016$ Ω, and $R^{N'} \approx 0.6 \pm 0.2$ Ω.

**Supplemental References**